\documentclass{aa}
\usepackage{graphicx}
\usepackage{txfonts}

\newcommand{\rms}{{\emph{rms}}}
\newcommand{\msun}{\ensuremath{M_\odot}}                          
\newcommand{\rsun}{\ensuremath{R_\odot}}                          
\newcommand{\vsini}{\ensuremath{\,v \sin i}}                  
\newcommand{\Msun}{\ensuremath{\,{\rm M}_\odot}}                  
\newcommand{\kms}{\,km\,s$^{-1}$}                                 
\newcommand{\mc}[1]{\multicolumn{2}{c}{#1}}                       
\newcommand{\sci}[2]{{:1}\!\times10^{:2}}                         

\newcommand{\etal}{et~al.}

\newcommand{\ie}{i.e.}
\newcommand{\eg}{e.g.}

\newcommand{\cf}{{cf.}}
\newcommand{\teff}{$T_{\rm eff}$}

\newcommand{\topp}{\emph{top}}

\newcommand{\bott}{\emph{bottom}}

\newcommand{\iotoph}{$\iota$~Oph}
\newcommand{\iotophh}{$\iota$~Ophiuchus}
\newcommand{\psicen}{$\psi$~Cen}
\newcommand{\psicenn}{$\psi$~Centauri}
\newcommand{\thetacen}{$\theta$~Cen}

\def\kms{\ifmmode{\rm km\thinspace s^{-1}}\else km\thinspace s$^{-1}$\fi}
\newcommand{\dss}{$\delta$~Scuti}

\newcommand{\str}{Str\"omgren}

\newcommand{\ebop}{{\sc ebop}}

\newcommand{\mad}{{\sc mad}}
\newcommand{\cles}{{\sc cl\'es}}
\newcommand{\wire}{{\sc wire}}
\newcommand{\smei}{{\sc smei}}
\newcommand{\hipparcos}{{\sc hipparcos}}
\newcommand{\vizier}{{\sc vizier}}
\newcommand{\simbad}{{\sc simbad}}
\newcommand{\kepler}{{\sc kepler}}
\newcommand{\corot}{{\sc corot}}

\begin{document}
      \title{Eclipsing binaries observed with the \wire\ satellite}
   \subtitle{I. Discovery and photometric analysis of the new bright A0\,IV eclipsing binary \psicenn}

   \author{H.\ Bruntt
          \inst{1,2}
          \and
          J.\ Southworth\inst{3}
          \and
          G.\ Torres\inst{4}
          \and
          A.J.\ Penny\inst{4}
          \and
          J.V.\ Clausen\inst{1}
          \and
          D.L.\ Buzasi\inst{5}
          }

   \offprints{H. Bruntt}

   \institute{
              Niels Bohr Institute, University of Copenhagen,
              Juliane Maries Vej 30, DK-2100 Copenhagen \O, Denmark
              \email{jvc@astro.ku.dk}
         \and
             {School of Physics A28, University of Sydney, 2006 NSW, Australia}
              \email{bruntt@physics.usyd.edu.au}
         \and
             Department of Physics, University of Warwick, Coventry, CV4 7AL, UK
             \email{jkt@astro.keele.ac.uk}
         \and
             Harvard-Smithsonian Center for Astrophysics,
             60 Garden Street, Cambridge, MA 02138, USA
             \email{gtorres@cfa.harvard.edu, apenny@cfa.harvard.edu}
         \and
             US Air Force Academy, Department of Physics, CO, USA
             \email{Derek.Buzasi@usafa.af.mil}
             }

   \date{Received xxx; accepted yyy}


  \abstract
{Determinations of stellar mass and radius with realistic uncertainties at the level of 1\% provide
important constraints on models of stellar structure and evolution.}
{We present a high-precision light curve of the A0\,IV star \psicenn, from the star 
tracker on board the \wire\ satellite and the Solar Mass
Ejection Imager camera on the Coriolis spacecraft. 
The data show that \psicen\ is an eccentric eclipsing binary 
system with a relatively long orbital period.}
{The \wire\ light curve extends over 28.7 nights and contains 41\,334 observations 
with 2\,mmag point-to-point scatter. The eclipse depths are 0.28 and 0.16\,mag, and
show that the two eclipsing components of \psicen\ 
have very different radii. As a consequence, the secondary eclipse is total.
We find the eccentricity to be $e=0.55$ with an orbital 
period of 38.8\,days from combining 
the \wire\ light curve with data taken over two years from the Solar Mass Ejection Imager camera.} 
{We have fitted the light curve with \ebop\ and have assessed the uncertainties of the resulting parameters 
using Monte Carlo simulations. 
The fractional radii of the stars and the inclination of the orbit have 
random errors of only 0.1\% and 0.01$^\circ$,
respectively, but the systematic uncertainty in these quantities may be somewhat 
larger.
We have used photometric calibrations to estimate the effective 
temperatures of the components of \psicen\ to be $10\,450\pm300$ and $8\,800\pm300$\,K
indicating masses of about 3.1 and 2.0\Msun. 
There is evidence in the \wire\ light curve for $g$-mode pulsations in the primary star.}
{}

   \maketitle
%

\section{Introduction}

The study of detached eclipsing binaries is of fundamental importance to stellar astronomy as a means by which we can measure accurately the parameters of normal stars from basic observational data (Andersen \cite{andersen91}). The masses and radii of the component stars of a detached eclipsing binary (dEB) can be measured to accuracies better than 1\% (e.g.\ Southworth et al., \cite{SMS05}), and the effective temperatures and luminosities can be obtained from spectral analysis or the use of photometric calibrations. An important use of these data is in the calibration of theoretical models of stellar evolution (Pols et al.\ \cite{pols+97}; Andersen et al.\ \cite{andersen1990}; Ribas et al.\ \cite{ribas+00}). Comparing the properties of a dEB to the predictions of theoretical models allows the age and metal abundance of the binary to be estimated (e.g.\ Southworth et al.\ \cite{SMSa}) and the accuracy of the predictions to be checked, particularly if the two components of the dEB are of quite different mass or evolutionary stage (e.g.\ Andersen et al.\ \cite{andersenetal1991}).

Some of the physical effects contained in current theoretical stellar evolutionary models are poorly understood, so are treated using simple parameterisations, for example convective core overshooting and convective efficiency (mixing length). Whilst these parameterisations provide a good way of including such physics, our incomplete knowledge of their parameter values is compromising the predictive power of the models (Chaboyer \cite{chaboyer95}, Young et al.\ \cite{young01}) and must be improved. The amount of convective core overshooting has been studied using dEBs by Andersen et al.\ (\cite{andersen1990}) and Ribas et al.\ (\cite{ribas+00}), and the mixing lengths in the binary AI~Ph\oe nicis has been determined by Ludwig \& Salaris (\cite{ludwig+99}), but these studies have not been able to provide definitive answers because the age and chemical composition are usually free parameters which can be adjusted when comparing the model predictions to the properties of the dEBs.

Progress in the calibration of convective overshooting and mixing length may be made by increasing the accuracy with which the basic parameters of dEBs are measured, which requires much improved observational data. It is arguably more difficult to significantly improve the quality and quantity of light curves, rather than radial velocity curves, because a large amount of telescope time is needed for each system. 
Forthcoming space missions such as \kepler\ (Basri et al.\ \cite{kepler05}) and \corot\ (Baglin et al.\ \cite{baglin01}) 
will help to solve this problem by obtaining accurate and extensive light curves of a significant number of dEBs.

We have begun a programme to obtain high-quality photometry of bright dEBs with the Wide field InfraRed Explorer (\wire) satellite, with the intention of measuring the radii of the component stars to high accuracy. Several of the targets in this program have long orbital periods (\psicen) or shallow eclipses (AR\,Cas; \cite{arcas}) which has made them difficult to observe using ground-based telescopes. \psicen\ and AR\,Cas have secondary components which have much smaller masses and radii than the primary stars, so will be able to provide strict constraints on the predictions of theoretical models. 
An additional advantage of having components with very different radii 
is that the eclipses are nearly total, 
which generally allows the radii to be measured with increased accuracy 
compared to systems with partial eclipses.

In this paper we report our discovery of eclipses in the bright ($V =
4.1$) A0\,IV star $\psi$~Cen made with the \wire\ satellite. We present a
detailed analysis of the light curve showing that the orbit is
eccentric and the secondary eclipse is total. As is turns out, this is a
particularly interesting system since its brightest component is
located in the region of the Hertzsprung-Russell diagram between the blue edge of the
Cepheid instability strip and the region of $g$-mode oscillations seen
in slowly pulsating B stars (see, e.g., Pamyatnykh \cite{pamyat1999}). We present
an analysis of the oscillations showing evidence of two low-frequency
modes that may be interpreted as global oscillation $g$-modes in the
primary star.


\section{Photometry from space: \wire, \smei\ and \hipparcos}

Since 1999 the star tracker on the \wire\ satellite (Hacking et al.\ \cite{hacking99}) 
has been used to observe around 250 stars which have apparent magnitudes $V<7$ 
(see Bruntt \etal\ \cite{bruntt05}). The aperture is 52\,$m$m and
data is acquired at a cadence of 2~Hz. 
Each star is monitored for typically 2--4 weeks
and some targets have been observed in more than one season.
The primary aim has so far been to map the variability
of stars of all spectral types, and recently we have done
coordinated ground- and space-based observations. 
When planning the \wire\ observations 
a primary target is chosen and in addition the on-board 
computer selects four additional bright stars within the 
field of view (FOV) of 8.5$^\circ$ by 8.5$^\circ$. 
The monitoring of five stars by the star tracker was originally 
designed to perform accurate attitude control. 
The \rms\ uncertainty
on the position is 0.0085 pixels or about 0.5 arc seconds.

When binning data to one point every 15 seconds
the \rms\ uncertainty per data point is 0.3 and 5\,mmag 
for a $V=0$ and $6$ star, respectively. 
The filter response of \wire\ is not well known. 
From an analysis of the light collected from stars 
of various spectral types and measured brightness we estimate the filter to 
approximate Johnson $V+R$. 
The amplitudes of variation of variable stars studied 
from the ground and with \wire\ are in rough agreement with this,
\eg\ \cite{bruntt_epscep} compared \str\ $y$ and \wire\ filter observations of
the same \dss\ star.

The orbital period of \wire, which has been slowly decreasing since its 
launch in 1999, was 93.8 minutes at the time of these observations. 
\wire\ switches between two targets during each orbit in order to best 
avoid the illuminated face of the Earth. Consequently, the duty cycle is 
typically around 20--30\% or about 19--28 minutes of continuous 
observations per orbit.

\begin{figure*}
   \centering
   \includegraphics[width=17.4cm]{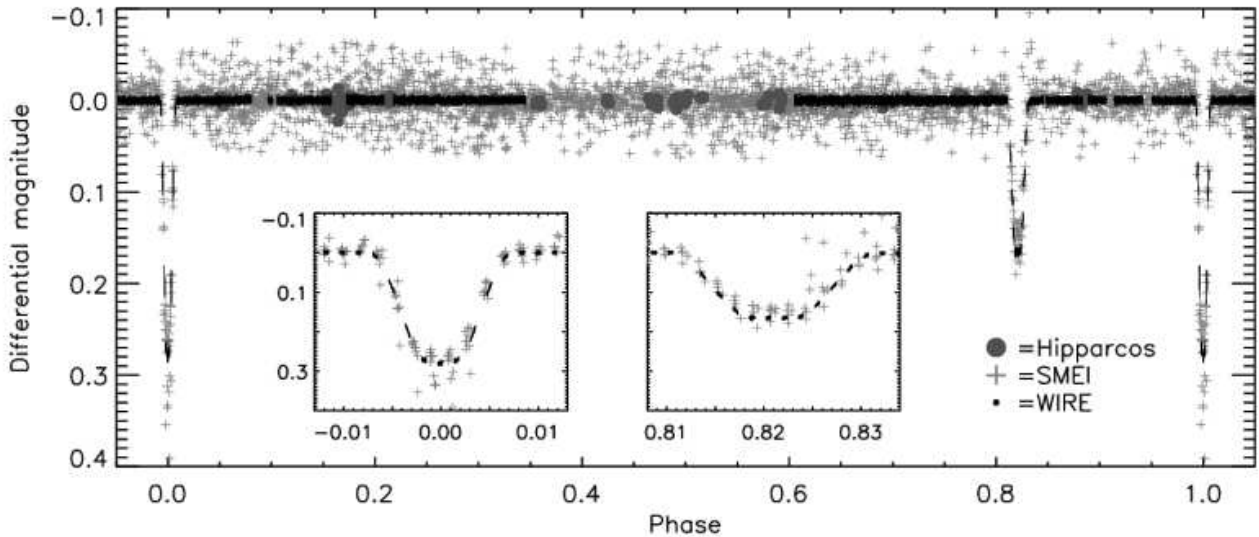}
      \caption{Phased light curve of \psicen\ for a period of 38.81252 days.
Data points from \wire\ (black dots), \smei\ (grey $+$ symbols),
and \hipparcos\ (grey circles) are shown.
              }
         \label{fig:ph}
   \end{figure*}

The A0\,IV type-star \psicen\ was observed nearly continuously for 28.7 days from 
June 11 to July 9 2004 as one of the secondary targets for the 
bright giant star \thetacen. 
After binning the light curve to a time sampling of 15 seconds we have 
41\,334 data points with a point-to-point scatter of only 2.1\,mmag. 
The light curve shows only one primary and one secondary eclipse separated in time by about 7 days, 
indicating that the orbit is eccentric and has a long period. 
The depths of the eclipses are about 0.28 mag for the primary and 0.16 mag for the secondary,
indicating that the components have quite different surface fluxes. 
The phased light curve from \wire\ is shown in Fig.~\ref{fig:ph}.


%
%

The \wire\ data only contain one primary and one secondary eclipse, 
so cannot be used to find the orbital period of \psicen. 
We therefore obtained data from 
the Solar Mass Ejection Imager (\smei) camera on the Coriolis satellite.
\smei\ is an all-sky camera aboard the U.S.\ Department of 
Defense Space Test program's Coriolis satellite which was 
launched on January 6 2003 into a 840 km high Sun-synchronous polar orbit. 
The primary purpose of \smei\ is to detect and track transient structures, 
such as coronal mass ejections emanating from the Sun by observing 
Thomson-scattered sunlight from heliospheric plasma (see Webb et al.\ \cite{webb06}). 
\cite{jackson04} provide the history of 
the \smei\ design and development, and the instrument design and testing 
are described by \cite{eyles04}. A description of its performance 
is given in Webb et al.\ (\cite{webb06}).


\smei\ consists of 3 CCD cameras each with an asymmetrical aperture of 1.76\,cm$^2$ area and
viewing a 3 x 60 degree strip of the sky, aligned end-to-end and slightly 
overlapping so as to provide a 3 degree wide strip extending 
approximately 160 degrees along a
great circle with one end near the direction of the Sun and the other at the
anti-Sun direction. The cameras are fixed on the zenith-nadir pointing 
Coriolis satellite, pointing some 30 degrees above the rear Earth horizon. 
Thus, during each 102 minute orbit of the satellite the cameras are swept 
to cover about 90\% of the entire sky, with a duty cycle of about 
85\% from launch until the present (March 2006). 
The unfiltered bandpass has a response which rises above 10\% between 
450 and 950 nm, with a peak response of about 50\%,
thus corresponding roughly to a wide $R$ photometric band. 
A G-type star with $V=10$ viewed on-axis produces approximately 800 photo-electrons 
during a single 4\,s exposure. Each camera takes an image every 
4 seconds (which are thus smeared roughly by 0.2 degrees {\em across} the 
picture). The camera pointing nearest the Sun (Camera~3) is used in a mode that 
gives it 
$0.1 \times 0.1$ degree binned pixels, whilst for the other 
two the binned pixels are $0.2 \times 0.2$ degrees.
The very fast (f/1.2$\times$f/2.2) optics are
used to give a point spread function (PSF), with a 
full-width at half-maximum of about 0.08 degrees but with wide wings. 
Whilst for the main project 
aim, the optical design is very suitable, for the aim of stellar photometry 
there are certain drawbacks, namely the short exposures, the optical distortion 
of the FOV and of the PSF, 
the large pixel size, and the variation 
of the CCD temperatures during an orbit. 
The main limitation, however, is the crowding.

We used \smei\ data covering two years from March 21 2003 to March 15 2005. 
Using only one of the three cameras 
(Camera 2) for these preliminary studies, the data have gaps varying 
between 87 and 140 days  when the star is visible only in one of the 
other two cameras.
The typical accuracy per data point is 
about 16\,mmag\footnote{The \smei\ data is capable 
of giving better accuracy. On short time scales an accuracy of 
about 1\,mmag is achieved, and the analysis method is being refined to achieve 
that accuracy on longer time scales.},  
which is significantly larger than the \wire\ accuracy of 2.1\,mmag. 
We have 3\,773 data points from \smei\ but a significant fraction are 
spurious outliers (see Sect.~\ref{sec:reduction}).     
The grey points in the phased light curve in Fig.~\ref{fig:ph} are the \smei\ data.

We also used 91 data points from \hipparcos\ (\cite{hip}) collected 
from February 1990 to February 1993.
Whilst none of these data points were taken during eclipse, they have 
been useful in helping to constrain the period of \psicen.

\begin{figure*} \resizebox{\hsize}{!}{\includegraphics{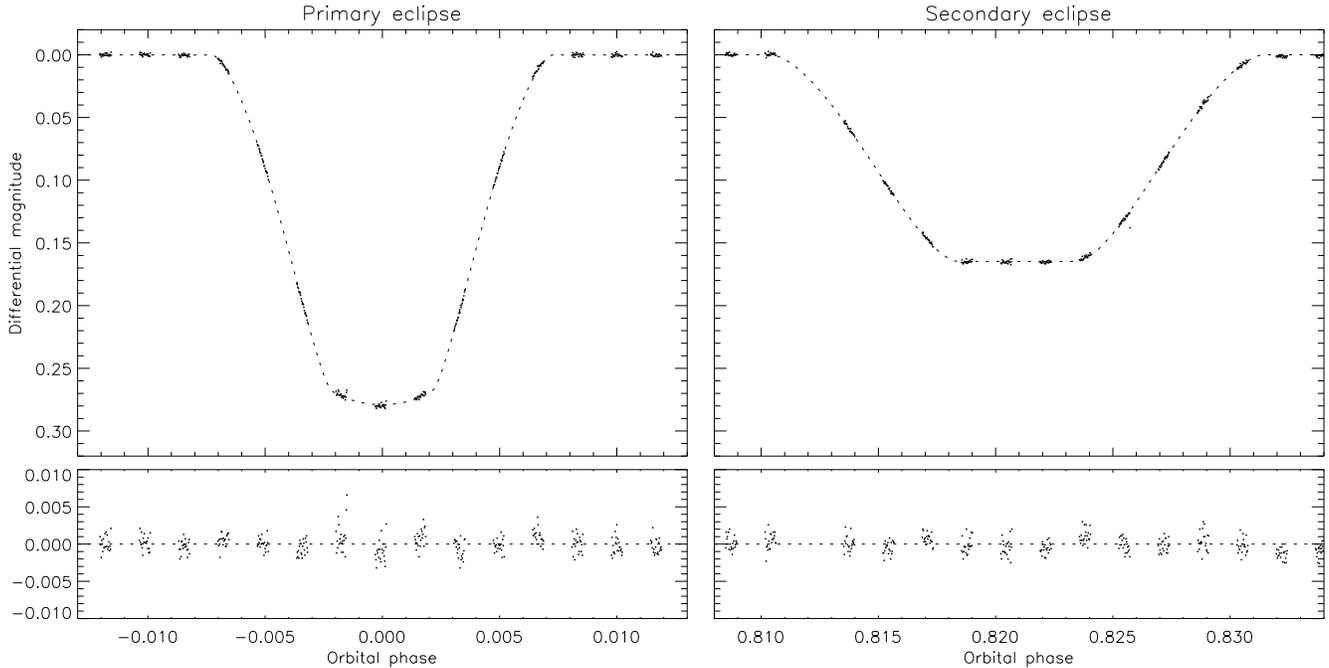}}
\caption{The \wire\ light curve during the
primary and secondary eclipses is compared to the best-fitting
{\sc ebop} model (dotted line). The residuals of the fit are shown below each panel
with an increased scale. In these plots groups of five points were binned together.}
\label{fig:lcfit} \end{figure*}

\subsection{Data reduction\label{sec:reduction}}

The \wire\ dataset consists of about 1.3 million windows 
extracted from the 512$\times$512 CCD star tracker camera.
Each window is 8$\times$8 pixels and aperture photometry is carried out. 
The basic data reduction was done using the pipeline described by 
Bruntt \etal\ (\cite{bruntt05}). 

In the \smei\ dataset \psicen\ will appear on approximately 30 out of the 20\,000 images from
each satellite orbit. For about 15 of these, the star will be adequately 
distant from the FOV edges (giving a total exposure in each orbit of about 
one minute). For each image used, the CCD data have bias and dark current removed. 
The dark flux for each pixel varies from exposure to exposure as the camera 
temperature changes. The significant variation in effective telescope aperture with off-axis 
direction is allowed for. Glare from scattered sunlight is
subtracted using an empirical map (Buffington, private communication, 2006). 
Further details of the glare are given in  \cite{buffington05}.
From the satellite pointing information each pixel in an approximately 
$1.2\times1.2$ degree area around the star is given an absolute sky direction, 
accurate to a few arc seconds. 
Then for all the binned pixels used (about 500) a PSF is fitted by a 
linearised least-squares method to derive the height and sloping base. 
The PSF is empirically determined from observations of Sirius taken when the 
cameras are being used in an optional $0.05 \times 0.05$ degree pixel high-resolution 
mode. 
Full details of the data reduction are planned to be 
given in Penny et al.\ (2006). Further discussion of stellar photometry 
with \smei\ is also given in \cite{buffington06}.

The \smei\ data contain a significant number of spurious outliers and for the subsequent 
analysis we removed these. Initially we used all data points to determine the approximate 
period. After phasing the data points with this period, we discarded data points that 
deviate more than 3.5\,$\sigma$ or 64\,mmag, but only for data points outside 
the times of eclipse. During times of eclipse we removed only five data points 
that deviated more than 150\,mmag from the mean phased light curve. 
A total of 325 spurious data points, or about $9\%$ of the \smei\ data, were discarded.

The \hipparcos\ photometry was extracted directly from the \vizier\ database. The 91 data points from \hipparcos\ have a mean magnitude of $H_{\rm p} = 4.039 \pm 0.004$. The mean magnitude outside eclipse was subtracted from each of the \wire, \smei\ and \hipparcos\ prior to subsequent analysis.

A small part of the \wire\ light curve is given
in Table~\ref{tab:lcdata} while the complete dataset from both \wire\ and \smei\ is
available in electronic format at Centre de Donn\'ees astronomiques de Strasbourg (CDS).

\begin{table} \caption{\wire\ and \smei\ observations of \psicen. The
magnitudes for each source have been normalised so the outside-eclipse
magnitude is zero. The full table comprises 44\,782 datapoints and 
is available at the CDS.}
\label{tab:lcdata} \centering
\begin{tabular}{l r r} \hline \hline
HJD $-$ 2400000 &  Magnitude  & Source \\ \hline
53167.7383428   &    0.000022 &  WIRE  \\
53167.7385222   & $-$0.000355 &  WIRE  \\
53167.7387016   &    0.002896 &  WIRE  \\
53167.7388847   & $-$0.000195 &  WIRE  \\
53167.7390662   & $-$0.000541 &  WIRE  \\
53167.7392467   &    0.000129 &  WIRE  \\
53167.7394309   & $-$0.001556 &  WIRE  \\
53167.7396159   &    0.000402 &  WIRE  \\
53167.7397979   & $-$0.001407 &  WIRE  \\
53167.7399805   & $-$0.001965 &  WIRE  \\
\hline \end{tabular} \end{table}

\section{Light curve analysis\label{sec:lc}}

The high quality of the \wire\ light curve of \psicen\ has allowed us to derive extremely precise photometric parameters for the stars. The orbital period of \psicen\ is not determined from the \wire\ data, which only contains one primary and one secondary eclipse for this eccentric system, so we have included the \smei\ data in our analysis as they contain observations during eclipses separated by up to two years.

Preliminary times of minimum light were derived from the \smei\ and \wire\ data by fitting Gaussian functions to the eclipses present in the data. An orbital period of 38.8119(9) days provides a good fit to these times of minima and the overall light curve of \psicen. The actual shape of the eclipses are however, quite non-Gaussian, and therefore the orbital period and the reference time of the central primary eclipse were included as fitting parameters in the light curve analysis. As seen in Table~\ref{tab:lcfit} we obtained a period of 38.81252(29) days. 
The phased light curve is shown in Fig.~\ref{fig:ph} and the
details of the eclipses seen with \wire\ are shown in Fig.~\ref{fig:lcfit}.

The \wire\ and \smei\ light curves were analysed with {\sc jktebop} (Southworth et al.\ \cite{SMSb}, \cite{SMSc}), which uses the NDE eclipsing binary model (Nelson \& Davis \cite{nelson72}) as implemented in the {\sc ebop} code (Etzel \cite{etzel81}; Popper \& Etzel \cite{popper81}). This model represents the projected shapes of the components of an eclipsing binary as biaxial spheroids, an approximation which is easily adequate for \psicen, whose components are almost spherical. A significant advantage of the model is that it involves very few calculations, meaning that extensive error analyses can be undertaken quickly using standard computing equipment. For this application, {\sc jktebop} was modified to allow the orbital period and time of primary mid-minimum to be included as free parameters.

Preliminary solutions of the \wire\ and \smei\ datasets were made using {\sc jktebop}, fitting for the orbital ephemeris and inclination, the sum and ratio of the fractional stellar radii, the linear limb darkening coefficients for each star, and the quantities $e\cos\omega$ and $e\sin\omega$ where $e$ is the orbital eccentricity and $\omega$ is the longitude of periastron. The gravity darkening exponents were fixed at 1.0 and the mass ratio at 0.65; large changes in these values have a negligible effect on the solution. Third light was fixed at zero because its value converged to a significantly negative (and so unphysical) value if it was included as a fitting parameter. The sizes of the residuals in the \smei\ data are 20\,mmag.
The residuals of the preliminary \wire\ light curve fit were analysed for asteroseismological signatures, and the two strongest modes were subtracted from the data (see Sect.~\ref{sec:intrinsic}). The final \wire\ light curve (41\,334 data points) has a \rms\ residual scatter of 2.1\,mmag. Once the observational uncertainties had been assessed for both the \wire\ and \smei\ data, we were able to combine them to obtain the final photometric solution. The slight difference in the wavelength dependence of the \wire\ and \smei\ data is unimportant here because the difference in quality and quantity between the two datasets means that the \smei\ data have a significant effect only on the derived orbital period. The photometric parameters found in this solution are given in Table~\ref{tab:lcfit} and the best fit to the light curve is plotted against the \wire\ data in Fig.~\ref{fig:lcfit}.

\begin{table}
\caption{Photometric parameters and uncertainties for
         \psicen\ found using {\sc jktebop} and Monte Carlo simulations.
          The first half of the table contains the fitted parameters and
          the second half the dependent quantities. 
          Note that 2\,453\,000\,days must be added to the time of the primary minimum, $T_0$. 
        \label{tab:lcfit}} 
\centering
\begin{tabular}{l c r@{\,$\pm$\,}l} \hline \hline
Parameter                                 &                                   &           \mc{Value}        \\
\hline
Fractional sum of the radii               & $r_{\rm A}+r_{\rm B}$             &     0.065861    & 0.000070  \\
Ratio of the radii                        & $k$                               &     0.49737     & 0.00054   \\
Orbital inclination (deg)                 & $i$                               &    88.955       & 0.012     \\
Surface brightness ratio                  & $J$                               &     0.688       & 0.011     \\
Primary limb darkening                    & $u_{\rm A}$                       &     0.256       & 0.009     \\
Secondary limb darkening                  & $u_{\rm B}$                       &     0.362       & 0.041     \\
$e\cos\omega$                             & $e\cos\omega$                     &     0.52035     & 0.00011   \\
$e\sin\omega$                             & $e\sin\omega$                     &     0.19036     & 0.00097   \\
Orbital period (days)                     & $P$                               &    38.81252     & 0.00029   \\
Primary minimum (HJD)                & $T_0$                             &   183.049310    & 0.000066  \\
\hline
Fractional primary radius                 & $r_{\rm A}$                       &     0.043984    & 0.000045  \\
Fractional secondary radius               & $r_{\rm B}$                       &     0.021877    & 0.000032  \\
Stellar light ratio                       & $\frac{L_{\rm B}}{L_{\rm A}}$     &     0.16348     & 0.00013   \\
Orbital eccentricity                      & $e$                               &     0.55408     & 0.00024   \\
Periastron longitude (deg)                & $\omega$                          &    20.095       & 0.098     \\
\hline \end{tabular}
\end{table}

\subsection{Light curve solution uncertainties}

The uncertainties in the photometric parameters were assessed using the Monte Carlo simulation algorithm contained in {\sc jkt\-ebop} (Southworth et al.\ \cite{SMSb}, \cite{SMSc}). In this algorithm, the best-fitting model light curve is evaluated at the times of the actual observations. Observational uncertainties are added and the resulting synthetic light curve is refitted, using initial parameter estimates which are perturbed versions of the best-fitting parameters of the real light curve. This is undertaken many times, and the uncertainties are derived from the spread in the best-fitting parameters for the synthetic light curves.

The \wire\ light curve was re-binned to reduce the number of data points by a factor of 5, and then the Monte Carlo algorithm was run 6\,000 times. The re-binning process makes a negligible difference to the best fit, but greatly reduces the amount of CPU time required for this analysis. The reduced $\chi^2$ of the best fit to the re-binned data is 1.15, indicating that there is a systematic contribution as well as a random contribution to the residuals of the fit\footnote{The equivalent number, before the subtraction of the two oscillation modes (\cf\ Sect.~\ref{sec:intrinsic}), was 1.56. This indicates that it was important to remove these modes before the final light curve analysis}. Inspection of the light curve in Fig.~\ref{fig:lcfit} suggests that there are slight offsets in the zero-points of some of the blocks of observations. The final uncertainties we quote for the light curve parameters found in the {\sc jktebop} analysis (Table~\ref{tab:lcfit}) are the standard deviations of the spread of parameter values from the Monte Carlo analysis multiplied by $\sqrt{1.15}$ to account for the small systematic uncertainties.

%
%
\begin{figure} \resizebox{\hsize}{!}{\includegraphics{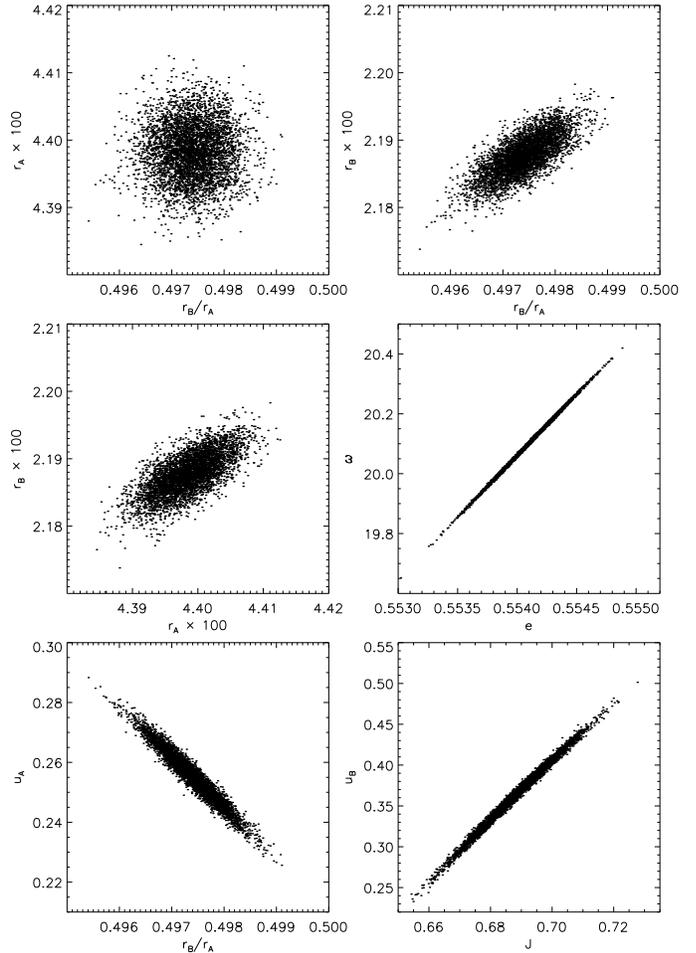}}
   \caption{Best-fitting parameter values for the 6\,000 synthetic
            light curves created for the Monte Carlo analysis.} \label{fig:MCplot}
\end{figure}
%

Fig.~\ref{fig:MCplot} shows the spread of the values of some best-fitting parameters for the synthetic light curves generated during the Monte Carlo analysis. The radii of the two stars are not strongly correlated with any other parameter or with each other, so their values are very reliable, but a strong correlation exists between eccentricity and periastron longitude; this effect has previously been noticed in CO\,Lac by \cite{wilson83}. 
However, the ranges of possible values for these and the other parameters are still very small and the correlation between eccentricity and periastron longitude has very little effect on the fractional radii of the stars.

As can be seen from Table~\ref{tab:lcfit}, the fractional radii of the component stars of $\psi$\,Cen are known to impressive random errors of 0.10\% (primary star) and 0.15\% (secondary), and all of the light curve parameters are extremely well determined. In particular, the linear limb darkening coefficient of the primary star is known to with $\pm$0.009, which to our knowledge is the most precise measurement ever in an eclipsing binary system. However, quoting such small random errors can be misleading because systematic errors, caused by imperfections in the model used to fit the light curve, may be somewhat larger (see below). Using the more complex logarithmic or square-root limb darkening laws may cause the radius measurements of the stars to increase about 0.2\% (Lacy et al.\ \cite{lacy05}).

Deriving uncertainties using the Monte Carlo algorithm in {\sc jktebop} makes the explicit assumption that the {\sc ebop} model is a good representation of the stars. The uncertainties in the photometric parameters arising from this assumption can be estimated by fitting the light curve with other eclipsing binary models, such as {\sc wink} (Wood \cite{wood71}) and the Wilson-Devinney code (Wilson \& Devinney \cite{wilson71}; Wilson \cite{wilson93}). We have attempted this and found large differences between the best fits for different light curve models. These differences have been traced to possible convergence problems with the differential corrections optimisation algorithms contained in the Wilson-Devinney code, {\sc wink}, and also the original {\sc ebop} code. We have avoided any convergence problems by generating model light curves, similar to that of $\psi$\,Cen, with the Wilson-Devinney code and using {\sc jktebop} to find a best fit ({\sc jktebop} uses the Levenberg-Marquadt minimisation algorithm as implemented in {\sc mrqmin} by Press et al.\ \cite{nr}). We find that the radii for the {\sc ebop} and Wilson-Devinney models agree to within 0.1\% (for a light curve with no eccentricity) and 0.3\% (for an eccentricity of 0.5), which is comparable to the random errors given in Table~\ref{tab:lcfit}. 
Further investigations into the true 
accuracy of the photometric parameters found from the \wire\ light curve will 
be deferred until we have modified the {\sc wink} and Wilson-Devinney codes by 
replacing the differential corrections subroutines with more modern algorithms.
We will then also be able to test more sophisticated limb darkening laws.

\section{Radial velocities\label{sec:radvel}}

Surprisingly little attention has been given to the measurement of the
radial velocity of such a bright star as $\psi$~Cen. The first
observations appear to be those made in 1914--1915 at the Lick
Observatory (\cite{Campbell1928}), but they show a large scatter even in
repeated measurements of the same three photographic plates. The
remark that the hydrogen lines are broad probably explains this. Four
additional measurements made in 1959--1960 from Mt.~Stromlo were
reported by \cite{Buscombe:61}. Those velocities also show a large
scatter, and do not phase up particularly well with the period and
epoch we have derived from the photometry. More recent measurements by
Grenier et al.\ (\cite{grenier:99}) are of higher quality (the formal uncertainty is $\sim$1~\kms)
and do agree well with the expected trend. No mention of double lines
has been made in the literature, although from our light-curve
solutions it is expected that the secondary will be visible in the
spectrum. A single-lined orbital solution based on the five
Grenier et al.\ \cite{grenier:99} velocities (attributed here to the primary star)
with the ephemeris, eccentricity, and longitude of periastron held
fixed at the photometric values gives a velocity semi-amplitude of 
$K = 37.4$~\kms, and a minimum companion mass of $M_2 \sin i = 0.495 \pm
0.011 (M_1 + M_2)^{2/3}$. However, reasonable estimates of the primary
mass ($\sim$3.1~M$_{\odot}$; see below) then lead to unreasonably
small values for the secondary ($\sim$1.3~M$_{\odot}$), which indicates
that the semi-amplitude $K$ is significantly underestimated, by
roughly a factor of 1.4. This is most likely due to line blending,
particularly given the large rotational broadening reported for the
star. Measured values of $v \sin i$ range from 100~\kms\ to 125~\kms\
(Slettebak et al.\ \cite{slettebak:75}, Grenier et al.\ \cite{grenier:99}, Royer et al.\ \cite{royer02}).

Absolute brightness measurements of $\psi$~Cen are available in a
number of photometric systems (\cite{Mermilliod:97}). We have
collected that information and used the measurements in the Johnson,
Str\"omgren, and Geneva systems along with colour-temperature
calibrations by
\cite{Popper80}, \cite{Moon85}, \cite{Gray92},
\cite{napi93}, \cite{Balona94}, \cite{Smalley95}, and
\cite{Kunzli97}
to estimate the mean effective temperature of the
binary. 
The result, 10\,200~K, along with our determination of the ratio
$J$ of the central surface brightness of the components leads to
individual temperatures of 10\,450~K and 8\,800~K, with estimated
uncertainties of 300~K. 
These correspond to approximate spectral types of B9 and A2.

We then used model isochrones from the Yonsei-Yale series
(\cite{Yi01, Demarque:04}) to infer the masses of the components by
seeking agreement with both temperatures and the ratio of the radii
for the system, under the assumption that the stars 
are coeval. For an assumed solar metallicity 
($Z = 0.01812$ in these models) we find a
good fit for an age of 290 Myr. The inferred masses are 3.1 and 2.0
\msun, with an estimated uncertainty of about 10\%. These mass estimates
are the basis for our claim above that the available velocities of
$\psi$~Cen give a biased value of the semi-amplitude $K$.


%
%
\section{Search for intrinsic variability in \psicen\label{sec:intrinsic}}

\begin{figure}
   \centering
   \includegraphics[width=8.8cm]{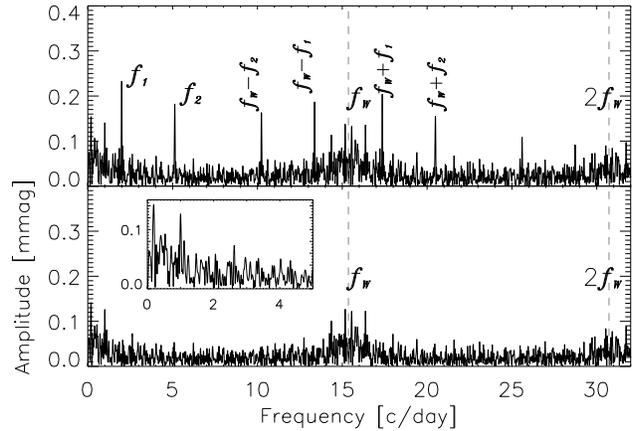}
      \caption{The amplitude spectrum of the \wire\ light curve of \psicen\ is
shown in the \topp\ panel. Two significant
peaks at $f_1$ and $f_2$ are found. 
The aliases and the \wire\ orbital frequency are marked.
The \bott\ panel is the amplitude spectrum after subtracting these two modes.
The inset shows the details at low frequencies; note the change in scale on
the ordinate.
              }
         \label{fig:amp}
   \end{figure}

Our estimate of the mass of the primary component of \psicen\
places it between the blue edge of the Cepheid instability
strip where \dss\ stars are found
and the region of slowly pulsating B-type stars (SPB) which
have masses above $\sim3~M_\odot$.
To search for intrinsic variability in \psicen\ we used
the \wire\ dataset and removed the data collected during the two eclipses.
We then calculated the amplitude spectrum
which is shown in the \topp\ panel in Fig.~\ref{fig:amp}.
There is a significant increase in the noise level towards low frequencies.
Due to the duty cycle of 30\% the signal at low frequencies
is leaked to the harmonics of the orbital frequency of \wire\ which is 
$f_W = 15.361\pm0.001$ c/day. 

We find two significant peaks at $f_1 = 1.996(2)$ and 
$f_2=5.127(3)$\,c/day with corresponding amplitudes of
0.23(1) and 0.18(1)\,mmag, and phases 0.65(6) and 0.82(8). 
This provides the best fit to $\Sigma_{i=1}^{2}a_i \sin [2\pi\,(f_i \cdot t+\phi_i)]$. 
The zero point in time for the phases is $T_0$ from Table~\ref{tab:lcfit}. 
The S/N levels of the modes are 6.0 and 8.2, respectively. 
We show the amplitude spectrum after
subtracting these two modes in the \bott\ panel in Fig.~\ref{fig:amp}.
The inset in the \bott\ panel shows an increase in the noise level
towards lower frequencies but none of the peaks below 5 c/day are significant.

To improve on this and look for additional variability
we subtracted the initial solution from the light curve, \ie\ the two 
modes $f_1$ and $f_2$ and the fitted binary light curve. We then
removed slow trends in the light curve by subtracting a sliding mean 
box with a width of 2 days (high-pass filtering). 
A slight dependence of the orbital phase and flux level was seen at the 
beginning of each orbit and we decorrelated this within the data.
The subtracted modes were then added to the improved light curve.
We find a significant change in the noise in the
light curve with time. During the first week and the last week
of observations the \rms\ noise level is about 40\% higher than in the
middle of the run. We assigned weights to each
data point, $i$, as $W_{{\rm time},i} \propto 1/\sigma_{\rm rms}^2$, 
where $\sigma_{\rm rms}$ is the \rms\ value within each group of 
350 data points (about three satellite orbits). 
These weights were then smoothed using a sliding box with a 
width of 1.3 days. Finally, we weighted outlier data points using 
$W_{{\rm outlier},i} \propto [1 + |\Delta m_i| / (a \cdot \sigma)^b]^{-1}$
with constants $a = 3.5$ and $b=8$, inspired by the discussion 
in point~5 in Sect.~4 of \cite{stetsonwei}.
Here $\Delta m_i$ is the differential magnitude of the $i$'th data point
and $\sigma$ is the \rms\ of the complete light curve.
The final weights are $W_{\rm time} \cdot W_{\rm outlier}$ which
are then normalized to unity.
Using the improved light curve and the weights described here 
we find the same frequencies and amplitudes of $f_1$ and $f_2$ 
within the uncertainties, but with the S/N improved by 20--30\%.

The primary aim of the procedure described here
was to improve the light curve 
by subtracting the intrinsic variation in the B-type component 
and also to improve the decorrelation with orbital phase
before making the final light curve analysis. This indeed 
improved the $\chi^2$ fit significantly as discussed in 
Sect.~\ref{sec:lc}.
The secondary aim of this procedure
was to detect additional low amplitude modes but none were found.

%
%

\subsection{Interpretation of $f_1$ and $f_2$\label{sec:interpret}}

The two detected peaks in the amplitude spectrum correspond to
periods of about 0.5 and 0.2 days. These periods could be
present in the data due to modulation caused by the rotation
period of the stars. To examine this possibility we must
estimate the {\em absolute} radii of the component stars.
To convert our accurate relative radii into absolute radii 
we estimate the semi-major axis from Kepler's third law with
the orbital period and the mass estimates given in Sect.~\ref{sec:radvel}.
The result is $3.7\pm0.2$\,\rsun\ for the primary 
and $1.8\pm0.1$\,\rsun\ for the secondary.

We use \vsini\ $=125\pm15$\,\kms\ for both the primary and secondary 
component to find rotational periods 
of $1.49\pm0.26$ and $0.74\pm0.13$\,days, respectively.
To reach a period as low as 
0.5 days the secondary star (with $R/\rsun = 1.8$) would need
to rotate with $v \sin i \simeq 180$~\kms. At present we cannot 
entirely rule this out, since the spectroscopically determined 
rotational velocity is dominated by the light from the primary.

To conclude, it seems difficult to explain the observed peaks in the
amplitude spectrum as due to the rotation of the primary and possibly
also the secondary. 

An alternate explanation is that the variation
is caused by high order low degree $g$-mode oscillations in the primary star.
The instability strip predicting the excitation of modes in
slowly pulsating B-type stars (SPB) was calculated for solar metallicity 
by Pamyatnykh (\cite{pamyat1999}).
However, we have \teff\ $=10,450\pm300$~K for the primary and this is on
the cool side of the theoretical SPB instability strip.

C.\ Aerts (private communication) kindly computed an evolution model 
for $M/M_\odot=3.1$ using the new solar composition
(\cite{asplund}) with the \cles\ code.
The pulsation code \mad\ was used to test the mode excitation 
(see, \eg\, \cite{dupret03} for more information on these codes).
None of the modes were found to be excited for this model for
\teff\ in the interval $10\,450\pm300$~K.
Ignoring this fact we compared the location of the two modes 
$f_1$ and $f_2$ to see if we could find a matching evolutionary 
state for modes with angular degree $l=1$ or $2$.

The modes in the pulsation model are closely spaced around the
low frequency mode $f_1$ and we cannot find an unambiguous fit. 
In the frequency range around $f_2$ the modes in the model are 
well separated but both $l=1$ and 2 are possible.
For $l=1$ there is a match of observed and computed frequencies 
at \teff\,$=10\,250$~K and for $l=2$ both \teff\,$=10\,200$~K and $10\,600$~K are possible.
However we must remember that these estimates are 
for solar metallicity while the metallicity of \psicen\ is unknown.
Accurate temperatures can be
found once photometry of the eclipses in multiple filters has been collected
and so a definite mode identification should be possible. 
Also, when the radial velocity curve of \psicen\ is acquired the masses should be
constrained to better than 1\% and a detailed modelling of this system
can be done. This should enable us to probe the extent of core
overshoot in the component stars.

It is important to note that
the model code does not take rotation into account.
\cite{aerts05} used the same evolution code to interpret observations
of the photometric variations seen in the main sequence SPB star HD~121190 which is
about 1\,500~K hotter than \psicen. 
They stress that
significant frequency shifts are expected for even moderately rotating
stars due to the Coriolis force perturbing the global oscillation modes
when the mode frequency is higher than half the rotation frequency. 
This is indeed the case for both modes seen in \psicen.
From the estimate made above we find $\Omega_{\rm rot} = 1/P_{\rm rot} = 0.67\pm0.12$\,c/day for 
the primary star when assuming $i_{\rm rot} \simeq i_{\rm orbit} = 89.0^\circ$.

Since current models do not predict oscillations at
the \teff\ of the primary star in \psicen\
another interesting possibility for the excitation
could be tidally induced modes.
\cite{willems02} investigated
this and found that in relatively close binary systems with highly
eccentric orbits it is possible for $l=2$
and $|m|=2$ modes to occur.
For this mechanism to work the pulsation 
frequency in the co-rotating frame must be an integer multiple of
the binary orbital frequency.
\psicen\ has a long period and so the orbital frequency is low,
$f_{\rm orbit} \simeq 0.026$\,c/day.
Hence, tidal interaction is most likely
not the excitation mechanism in this system.
Another explanation for the observed modes 
is excitation of retrograde mixed modes as
suggested by \cite{townsend05}.
To explore these possibilities further would require
detailed modelling of the \psicen\ system, which is
beyond the scope of the current study.

We should mention that \wire\ data are available
for the somewhat hotter star \iotophh\ (HD~152614).
The spectral type is B8V and we find \teff\ $=12\,000$~K from
\str\ photometry using the calibration by \cite{napi93}.
This star is within the SPB instability strip
predicted by Pamyatnykh (\cite{pamyat1999}). 
The amplitude spectrum shows at least
four significant modes at 0.82, 2.0, 4.6 and 4.9 c/day.
The amplitudes are 0.5\,mmag for the low frequency mode 
and $\simeq0.2$\,mmag for the remaining three modes. 
Thus, the frequency range and amplitudes are similar to \psicen.
We obtained a \cles\ model for this star but the \mad\ oscillation
code does not predict any modes above 1.5 c/day to be excited
for angular degrees $l=1$ and $2$.

Lastly, we see from Fig~\ref{fig:amp} that in the range 
of \dss\ star variation with periods of $\simeq$1--2 hours
(\ie\ frequencies from 12--24 c/day) no significant 
peaks are present with amplitudes above 0.1~mmag. 
In the region 14--17 c/day this claim is not valid
due to the increased noise level from the aliasing.
From the light curve solution we have $L_B/L_A\simeq0.163$ and 
so no \dss\ pulsation with amplitudes above $\simeq 0.6$\,mmag 
is present in the secondary component.

\section{Conclusions}

We report the discovery that \psicen\ is a bright detached eclipsing binary. 
Photometry from the star tracker on the \wire\ satellite shows that the orbit 
is eccentric and the period longer than about a month. With the photometry 
from the \smei\ camera we have been able to determine 
the period of the system to be 38.81252(29)\,d.

We have fitted the \wire\ and \smei\ light curves using {\sc ebop} to find 
precise photometric parameters for \psicen. Monte Carlo simulations show that 
the random error in the fractional radii of the stars is about 0.1\%.
We have attempted to assess the systematic uncertainties in this modelling 
by obtaining solutions with the {\sc wink} and Wilson-Devinney codes as well 
as with {\sc ebop}, but have so far been unsuccessful. Experiments have 
shown that this is probably due to convergence problems with the differential 
correction algorithms in the comparison light curve codes. We have been able 
to find that the systematic uncertainty in the radii are at most 0.3\%, and a more 
accurate assessment can be made once the comparison light curve codes have 
been modified to use a different optimisation algorithm.

Only a very limited number of radial velocity measurements for \psicen\ are 
available in the literature, and are severely affected by line blending,
so new high-resolution spectroscopy will be acquired 
by our group. With this data we will be able to measure the absolute masses and radii 
of the component stars of \psicen\ with high accuracy, which should allow us 
to place strong constraints on the predictions of theoretical stellar models 
for masses around 2--3\Msun. This will be helped if we are able to determine 
a precise metal abundance from the spectra of \psicen, but the high rotational 
velocity of the primary component may make this difficult.

We found two significant peaks in the amplitude spectrum
at 2.0 and 5.1 c/day with amplitudes of only 0.2~mmag. 
These frequencies are too high to be due to rotational modulation. 
We tentatively interpret them as low amplitude $g$-mode oscillations in the primary star.
A preliminary comparison with pulsation models shows that
a mode identification should be possible
once \teff\ and the mass of primary is in hand.
Thus, \psicen\ will be an interesting object both
as a classical detached eclipsing binary system and as an object for a detailed
asteroseismic study. It is very interesting that our results
indicate that the region of SPB stars as
determined from theory (Pamyatnykh \cite{pamyat1999}) may
have to be extended towards the blue edge of the
Cepheid instability strip for main sequence stars.
For further discussion of this point we refer
to \cite{aerts05}.

%
\begin{acknowledgements}
\smei\ is a collaborative project of the U.S. Air Force Laboratory, NASA, the
University of California at San Diego, and the University of Birmingham, U.K.,
which all have provided financial support. 
Details about the \smei\ instrument can be found at: 
{\tt http://www.vs.afrl.af.mil/factsheets/SMEI.swf} 
in addition to archival images, movies, and presentations at
{\tt http://smei.nso.edu}. The work of AJP was supported under 
NASA Research Grant NNG05GA41G to the SETI Institute.

The project {\em stellar structure and evolution --
new challenges from ground and space observations} is carried
out at Aarhus University and Copenhagen University and is supported
by the Danish Science Research Council
({\em Forskningsr\aa det for Natur og Univers}).
HB is supported by the Danish Science Research Agency
and HB and JS were both supported by the 
Instrument center for Danish Astrophysics (IDA)
in the form of postdoctoral grants.
JS would like to thank Tom Marsh and Boris G\"ansicke for many useful discussions.

GT acknowledges partial support for this work from the US National
Science Foundation grant AST-0406183 and NASA's MASSIF SIM Key Project (BLF57-04). 

We thank Conny Aerts for providing
\cles\ evolution models and pulsation predictions 
of \psicen\ and for many useful suggestions.

The following internet-based resources were used in research for
this paper: the NASA Astrophysics Data System, the \simbad\ database
and the \vizier\ service operated by {\sc cds}, Strasbourg, France, and the
ar$\chi$iv scientific paper preprint service operated by Cornell University.
\end{acknowledgements}

%


\begin{thebibliography}{}


\bibitem[Aerts \& Kolenberg (2005)]{aerts05} Aerts, C., \& Kolenberg, K.\ 
 2005, \aap, 431, 615

\bibitem[1990]{andersen1990}   Andersen, J., Clausen, J.~V., Nordstr\" om, B.\
 1990, \apj, 363, L33

\bibitem[1991]{andersen91}    Andersen, J.\
 1991, \aapr, 3, 91

\bibitem[1991]{andersenetal1991}   Andersen, J., Clausen, J.~V., Nordstr\" om, B., Tomkin, J., Mayor, M.\
 1991, \aap, 246, 99

\bibitem[Asplund et al.\ 2004]{asplund} Asplund, M., Grevesse,
  N., Sauval, A.~J., Allende Prieto, C.,  Kiselman, D.\  
 2004, \aap, 417,                               
 751, erratum:  Asplund et al.\ 2005, \aap, 435, 339

 \bibitem[2001]{baglin01} Baglin, A., Auvergne, 
   M., Catala, C., Michel, E., \& \corot\ Team 
 2001, ESA SP-464: SOHO 10/GONG 2000 Workshop: 
 Helio- and Asteroseismology at the Dawn of the Millennium, 10, 395 

\bibitem[Balona (1994)]{Balona94} Balona, L.~A.\
 1994, \mnras, 268, 119

\bibitem[2005]{kepler05} Basri, G., Borucki, W.~J.,  Koch, D.\ 
 2005, New Astronomy Review, 49, 478 

\bibitem[2005]{breger05} Breger, M. et al.\
 2005, \aap, 435, 955

\bibitem[2005]{bruntt05} Bruntt, H., Kjeldsen, H., Buzasi, D.~L., Bedding, T.\
 2005, \apj, 633, 440

\bibitem[Bruntt et al.\ (2006)]{bruntt_epscep} Bruntt, H., Su\'arez, J.~C., Buzasi, D.~L.\ et al.\
 2006, \aap, {\em submitted} 

\bibitem[Buffington et al.\ (2005)]{buffington05} Buffington, A., Jackson, B.~V., Hick, P.~P.\  
 2005, in Solar Physics and Space Weather Instrumentation, Proc.\ of SPIE, Vol. 5901, 590118, 
       doi: 10.1117/12615526

\bibitem[Buffington et al.\ (2006)]{buffington06} Buffington, A., Band, D.~L., 
          Jackson, B.~V., Hick, P.~P., Smith, A.~C.\
 2006, \apj,  637, 880

\bibitem[Buscombe \& Morris (1961)]{Buscombe:61} Buscombe, W., \& Morris, P.~M.\
 1961, \mnras, 123, 233

\bibitem[Campbell 1928]{Campbell1928} Campbell, W.~W.\
 1928, Publ.\ Lick Obs., 16, 1

\bibitem[1995]{chaboyer95}    Chaboyer, B.\
 1995, \apj, 444, L9

\bibitem[Demarque et al.\ 2004]{Demarque:04} Demarque, P., Woo, J.-H.,
  Kim, Y.-C.,  Yi, S.~K.\
 2004, \apjs, 155, 667

 \bibitem[Dupret et al.\ 2003]{dupret03} Dupret, M.-A., 
  De~Ridder, J., De Cat, P., Aerts, C., Scuflaire, R., Noels, A.,  Thoul, A.\
 2003, \aap, 398, 677


\bibitem[ESA 1997]{hip} ESA 
 1997, The \hipparcos\ and {\sc tycho} Catalogues, ESA SP-1200,
  \vizier\ Online Data Catalog, 1239

\bibitem[1981]{etzel81}       Etzel, P.~B.\
  1981,
  in eds.\ E.~B.\ Carling and Z.\ Kopal, Photometric and
  Spectroscopic Binary Systems, NATO ASI Ser.\ C., 69, Dordrecht, p.\,111

\bibitem[Eyles et al.\ (2003)]{eyles04} Eyles, C.~J.\ et al.\ 
  2003, \solphys, 217, 319 

\bibitem[Gray (1992)]{Gray92} Gray, D.~F.\
 1992, The Observation and Analysis of Stellar
  Photospheres, 2nd Ed. (Cambridge: Cambridge Univ.\ Press), 430

\bibitem[1999]{grenier:99}
  Grenier, S., Burnage, R., Faraggiana, R., Gerbaldi, M., Delmas, F.\
   Gomez A.~E., Sabas, V.,  Sharif, L.\
  1999, \aaps, 135, 503


\bibitem[1999]{hacking99} Hacking, P. et al.\ 
 1999, ASP Conf.~Ser.~177: Astrophysics with Infrared Surveys: A Prelude to 
  SIRTF, 177, 409 

\bibitem[Jackson et al.\ (2004)]{jackson04} Jackson, B.~V.\ et al.\ 
 2004, \solphys, 225, 177 

\bibitem[K\"unzli et al.\ (1997)]{Kunzli97}
  K\"unzli, M., North, P., Kurucz, R.~L.,  Nicolet, B.\
 1997, \aaps, 122, 51

\bibitem[2005]{lacy05}        Lacy, C.~H.~S., Torres, G.\ 
  Claret, A., Vaz, L.~P.~R.\
 2005, \aj, 130, 2838

\bibitem[1999]{ludwig+99}     Ludwig, H.-G., Salaris, M.\
 1999, in ASP Conf.\ Ser.\ 173, 
  Stellar Structure: Theory and Test of Connective Energy Transport, 
  eds.\ \'A. Gim\'enez, E. F. Giunan, \& B. Montesinos, 229

\bibitem[Mermilliod et al.\ 1997]{Mermilliod:97}
 Mermilliod, J.-C., Mermilliod, M.,  Hauck, B.\
 1997, \aaps, 124, 349

\bibitem[Moon \& Dworetsky (1985)]{Moon85}
 Moon, T.~T., \& Dworetsky, M.~M.\
 1985, \mnras, 217, 305

\bibitem[Napiwotzki et al.\ (1993)]{napi93} Napiwotzki, R.\
  Schoenberner, D.,  Wenske, V.\ 
 1993, \aap, 268, 653 

\bibitem[1972]{nelson72}      Nelson, B., Davis, W.~D.\
 1972, \apj, 174, 617

\bibitem[1999]{pamyat1999} Pamyatnykh, A.~A.\
 1999, Act.\ Astron., 49, 119

\bibitem[Penny et al.\ (2006)]{penny06} Penny, A.~J., Mizuno, D.~R., Buffington, A.\
 2006, ApJ, {\em in preparation}

\bibitem[1997]{pols+97}       Pols, O.~R.;, Tout, C.~A., Schr\"oder, 
  K.-P., Eggleton, P.~P., Manners, J.\
 1997, \mnras, 289, 869

\bibitem[Popper (1980)]{Popper80} Popper, D.~M.\
 1980, \araa, 18, 115

\bibitem[1981]{popper81}      Popper, D.~M., Etzel, P.~B.\
 1981, \aj, 86, 102

\bibitem[1992]{nr}           Press W.~H., Teukolsky S.~A., Vetterling, W.~T., Flannery B.~P.\
  1992,
  Numerical Recipes in Fortran 77: The Art of Scientific Computing,
  Cambridge University Press, p.\ 402.

\bibitem[2000]{ribas+00}      Ribas, I., Jordi, C., Gim\'enez, \'A.\
 2000, \mnras, 318, L55

\bibitem[2002]{royer02}
 Royer, F., Grenier, S., Baylac, M.-O., Gomez, A.~E.,  Zorec, J.\
 2002, \aap, 393, 897

\bibitem[1975]{slettebak:75} Slettebak, A., Collins, G.~W.~II, Boyce, P.~B., White, N.~M.,  Parkinson, T.~D.\
 1975, \apjs, 29, 137

\bibitem[Smalley \& Dworetsky (1995)]{Smalley95} Smalley, B., \& Dworetsky, M.~M.\
 1995, \aap, 293, 446

\bibitem[2004a]{SMSa}         Southworth, J., Maxted, P.~F.~L., Smalley, B.\
 2004a, \mnras, 349, 547

\bibitem[2004b]{SMSb}         Southworth, J., Maxted, P.~F.~L., Smalley, B.\
 2004b, \mnras, 351, 1277

\bibitem[2004c]{SMSc}         Southworth, J., Zucker, S., Maxted, P. F. L., Smalley, B.\
 2004c, \mnras, 355, 986

\bibitem[2005]{SMS05}         Southworth, J., Maxted, P.~F.~L., Smalley, B., Claret, A., Etzel., P.~B.\
 2005, \mnras, 363, 529

\bibitem[Southworth et al.\ 2006]{arcas}  Southworth, J., Bruntt, H., Busazi, D.~L.\
 2006, \aap, {\em in preparation}

\bibitem[Stetson (1990)]{stetsonwei} Stetson, P.~B.\ 
 1990, \pasp, 102, 932   

\bibitem[Townsend (2005)]{townsend05} Townsend, R.~H.~D.\ 
 2005, \mnras, 364, 573 
 
\bibitem[2006]{webb06} Webb, D.~F.\ et al.\ 
 2006, J.\ Geophys.\ Res., {\em submitted}

\bibitem[Willems \& Aerts (2002)]{willems02} Willems, B., \& Aerts, C.\ 
 2002, \aap, 384, 441

\bibitem[1971]{wilson71}      Wilson, R.~E., Devinney, E.~J.\
 1971, \apj, 166, 605

\bibitem[Wilson \& Woodward (1983)]{wilson83} Wilson, R.~E., \& Woodward, E.~J.\ 
 1983, \apss, 89, 5 

\bibitem[1993]{wilson93}      Wilson, R.~E.\
 1993, in ASP Conf.\ Ser.\ Vol.\ 38,
  New Frontiers in Binary Star Research, eds. Leung, K.-C., Nha, I.-S., p.\ 91

\bibitem[1971]{wood71}        Wood, D.~B.\
 1971, \aj, 76, 701

\bibitem[Yi et al.\ 2001]{Yi01} Yi, S.~K., Demarque, P., Kim, Y.-C.,
 Lee, Y.-W., Ree, C.~H., Lejeune, T.,  Barnes, S.\
 2001, \apjs, 136, 417

\bibitem[2001]{young01} Young, P.~A., Mamajek, 
 E.~E., Arnett, D.,  Liebert, J.\
 2001, \apj, 556, 230 


\end{thebibliography}
\end{document}